\documentclass[prl, twocolumn]{revtex4}

\usepackage{graphicx}
\usepackage{xcolor, amsmath}

\begin{document}

\title{Dynamical Exclusion of Probability and Energy Conservation}

\author{Victor Atanasov}
\address{Faculty of Physics, Sofia University, 5 blvd J. Bourchier, Sofia 1164, Bulgaria}
\email{vatanaso@phys.uni-sofia.bg}

\begin{abstract}
The interrelationship between energy and probability conservation is explored from the point of view of statistical physics and non-relativistic quantum mechanics. The simultaneous validity of the law of conservation of energy and the continuity equation (probability conservation) breaks for an interacting dynamical system. A separate and independent description of a physical system can be obtained by requiring that the law of conservation of probability is at the heart of the derivation of the ''equations of motion''. In effect, The Schr\"odinger equation can be viewed as an appropriate factorization of the continuity equation instead of an energy conservation relation per se. 
\end{abstract}

\maketitle

Physics is founded on the premise that some physical quantity is conserved \cite{goldstein}. Originally, Newton suggested that to every action, there is an equal and opposite reaction which metaphysical statement leads to the laws of conservation of energy and momentum in a closed system. Following the Newtonian line of thought we arrive at fluid dynamics, where the continuity equation arises as a statement of the mass conservation \cite{batchelor}. Now we know that mass conservation holds true only in a non-relativistic contex, but this does not render the continuity equation useless.

On the contrary, as an epiphenomenon of the existence and conservation of electrical charge, the continuity equation re-emerges as a mathematical expression of the statement that charge is neither created nor destroyed in an arbitrary volume. This law stands in Physics as a postulate emerging from experimental evidence \cite{EM}. 

Later on, Quantum Mechanics re-invented the continuity equation as a consequence of the quantum dynamics (Schrodinger or Dirac equations). However, this time around the continuity equation starts a life as a statement of the continuity of the probability density flux. In effect, one can read it as {\it probability is neither created nor destroyed} but can flow as a fluid. Probability conservation emerges as the dynamical restriction behind reality.

Naturally, we see that the law of conservation of energy and the continuity equation (the law of conservation of probability) are distinctly different statements about a physical system. Here we address the questions: Are these two distinct statements compatible? Do they hold simultaneously?

\section{An Observation in Statistical Physics}

Consider a 2-level system (ensemble of such systems) with $E_1$ and $E_2$ levels occupied by $N$ number of particles.  The number of particles at each state is $N_1$ or $N_2,$ therefore the probability of finding a particle at each state is $n_1=N_1/N$ and $n_2=N_2/N.$ Here
\begin{eqnarray}\label{consProb}
n_1+n_2=1    
\end{eqnarray}    
can be interpreted as either the law of conservation of probability or particle number.
In the case of a dynamical process, that is the number of particles at each state changes with time ($d/dt =\dot{}\;$), the following holds as a function of the conservation of probability
\begin{eqnarray}\label{dotProb}
\dot{n}_1+\dot{n}_2=0.    
\end{eqnarray}    

First, let us consider the classical Maxwell-Boltzman-Gibbs statistics, namely
\begin{eqnarray}\label{}
n_1=\alpha e^{-\frac{E_1}{k_B T}}    &&
n_2=\alpha e^{-\frac{E_2}{k_B T}}    
\end{eqnarray}    
Here $k_B$ is Boltzmann's consant, $T$ is temperature and $\alpha=\left[e^{-\frac{E_1}{k_B T}} + e^{-\frac{E_2}{k_B T}}\right]^{-1}$ is a normalization factor by virtue of (\ref{consProb}). 
We solve these relations for the energy levels and upon symmetrization of the $\Delta {E}=E_1-E_2$ difference we obtain
\begin{eqnarray}\label{}
\frac{\Delta {E}}{k_B T}&=&\frac12 \left[ \ln{\left(\frac{1}{n_1} - 1  \right)} - \ln{\left(\frac{1}{n_2} - 1  \right)} \right]. 
\end{eqnarray}    
Let us now differentiate with respect to time (keeping embedding temperature constant) to get
\begin{eqnarray}\label{}
\Delta \dot{E} \neq 0.
\end{eqnarray}    
The energy difference is not conserved, while the probability is. Obviously, if the probability is non-dynamic, i.e. static system, then both conservation laws hold simultaneously. 

Next, let us consider the Bose-Einstein/Fermi-Dirac statistics for the same system
\begin{eqnarray}\label{}
n_1=\frac{1}{e^{\frac{E_1}{k_B T}} - \epsilon}    &&
n_2=\frac{1}{e^{\frac{E_2}{k_B T}} - \epsilon} 
\end{eqnarray}    
where $\epsilon=1$ corresponds to the Bose-Einstein case and $\epsilon=-1$ to the Fermi-Dirac case.
Note, here that instead of (\ref{consProb}) we will use 
$n_1+n_2=a,$ where $a$ is a constant, therefore (\ref{dotProb}) still holds.   
$\Delta {E}$ is given by
\begin{eqnarray}\label{}
\frac{\Delta {E}}{k_B T}&=&\left[ \ln{\left(\frac{1}{n_1} +\epsilon  \right)} - \ln{\left(\frac{1}{n_2} +\epsilon  \right)} \right] 
\end{eqnarray}    
and when one differentiates with respect to time, while keeping embedding temperature constant) one obtains
\begin{eqnarray}\label{}
\Delta \dot{E} \neq 0.
\end{eqnarray}    
for the bosonic and the fermionic case. Naturally, the same reasoning holds: provided the probability is conserved we cannot expect energy conservation in the dynamic case.

Probability and Energy can be conserved simultaneously in the case $\dot{n}_i=0,$ where $i={1,2}$, which is considered a stationary state in the quantum mechanical setting. 

The result of this analysis, amongst many things, points to an arising question on the meaning of the governing equations in quantum mechanics: {\it Are they energy or probability conservation relations?} In the stationary case, where $d |\psi|^2/dt=0$ these equations are energy and probability conservation equations simultaneously, while in the dynamic case, they are not.

\section{Quantum mechanical case}

Let us take the non-relativistic quantum mechanical description, that is the Schr\"odinger equation, and explore the mutual dynamic exclusion of probability and energy conservation. 

The Schr\"odinger equation can be interpreted as an energy conservation relation due to its dimensionality: Joules on the left equal Joules on the right
\[
i \hbar \frac{\partial}{\partial t} \psi \; [J ]= \hat{H} \psi \; [J ].
\]
This equation incorporates in itself the conservation of probability through the continuity equation applicable to the probability density $|\psi |^2$ and probability current density $\vec{j}$. Here we explore this deeply engrained premise to reveal that it holds only in the static case. Moreover, the fact that the Schr\"odinger equation leads to the continuity equation in either dynamical or static case, point to the quantum mechanical description actually being a proper factorization of the continuity equation, rather than a product of energy conservation.

\subsection{Time-dependent potential}

We take the Schrodinger equation with a time-varying potential $V(t)$: $
i \hbar \frac{\partial \psi}{\partial t} = - \frac{\hbar^2}{2m} \Delta \psi + V(t) \psi(t,\vec{r})$.
Let us solve for $\psi(t,\vec{r}) = \chi(t)\phi(\vec{r})$ to split the dependance on the variables in two equations
\begin{eqnarray}\label{eq_chi1}
i \hbar \partial_t \chi - V(t)\chi &=& E \chi(t)\\\label{eq_phi1}
- \frac{\hbar^2}{2m} \Delta \phi &=& E \phi(\vec{r})
\end{eqnarray}    
Equation (\ref{eq_chi1}) has a solution 
\begin{eqnarray}\label{chi1}
\chi(t) &=& \chi_0 e^{-i \frac{Et}{\hbar}} e^{\frac{-i}{\hbar} \int^{t} {V(t')dt'} }
\end{eqnarray}    
while equation (\ref{eq_phi1}) is the standard Helmholtz equation which we solve in the one-dimensional case where it converts to the oscillator equation
$d^2 \phi / dx^2 + \omega^2 \phi$. Here $\omega^2 = 2mE/\hbar^2$ and one linearly independent solution is given by
\begin{eqnarray}\label{phi1}
\phi(x) &=& \phi_0 \sin{(\omega x)}
\end{eqnarray}    

Let us check whether probability is conserved:$
| \psi(t,\vec{r})|^2 = \chi_0^2  [\phi_0 \sin{(\omega x)} + \phi_1 \cos{(\omega x)}]^2 ,$
therefore
\begin{eqnarray}
\partial_t | \psi(t,\vec{r})|^2 &=& 0.
\end{eqnarray}    
The current density 
\begin{eqnarray}\label{current j}
\vec{j}=-\frac{i\hbar}{2m} (\psi^{\ast} \nabla \psi - \psi \nabla \psi^{\ast})
\end{eqnarray}    
for the above one-dimensional solution (\ref{chi1}) and (\ref{phi1}) yields:
\begin{eqnarray}
j_x=-\frac{i\hbar}{2m} (\psi^{\ast} \frac{d \psi}{d x} - \psi \frac{d \psi^{\ast}}{dx}) =0\\
j_y=j_z=0
\end{eqnarray}    
As a result the conservation of probability $\partial_t | \psi|^2+ {\rm div} \vec{j}=0$ holds. However, energy is non-conserved. Let us evaluate the energy of that state $
E_{{\rm state}}= i \hbar \int \psi^{\ast} \frac{\partial \psi}{\partial t} dr^3 = E + V(t), $    
which turns out to be time-dependent and
\begin{eqnarray}
\frac{ d E_{{\rm state}} }{d t} \neq 0
\end{eqnarray}    
energy is not conserved.

We can summarise the case by the observation that when the non-relativistic quantum system is subject to a time-dependent potential we have conservation of probability but not conservation of energy. Therefore, the Schr\"odinger equation can be considered a peculiar form of factorisation of the continuity equation, rather than an energy conservation statement.

\subsection{Space-dependent potential}

Take the Schrodinger equation with a space-varying potential $V(\vec{r})$: $i \hbar \frac{\partial \psi}{\partial t} = - \frac{\hbar^2}{2m} \Delta \psi + V(\vec{r}) \psi(t,\vec{r})
$. Let us solve for $\psi(t \vec{r}) = \chi(t)\phi(\vec{r})$ to obtain
\begin{eqnarray}\label{eq_chi2}
i \hbar \partial_t \chi &=& E \chi(t)\\\label{eq_phi2}
- \frac{\hbar^2}{2m} \Delta \phi + V(\vec{r})\phi &=& E \phi(\vec{r})
\end{eqnarray}    
The solution to (\ref{eq_chi2}) is a straight-forward one
\begin{eqnarray}\label{chi2}
\chi(t) &=& \chi_0 e^{-i \frac{Et}{\hbar}} ,
\end{eqnarray}    
while an exact solution to (\ref{eq_phi2}) does not exist in the general case. We will seek an approximate one. Let us first rewrite the equation in the following form
\begin{eqnarray}\label{eq_phi2_2}
\Delta \phi + k^2(\vec{r}) \phi &=& 0,
\end{eqnarray}    
where $k^2(\vec{r}) = \frac{2m}{\hbar^2}  [E - V(\vec{r})]$. Here we can see that in the case of a periodic potential, the periodicity is translated into $k$ and therefore a model solution similar to Kronig-Penney case has a potential to reveal the underlying behaviour. 

We can further simplify by assuming a one-dimensional motion, therefore equation (\ref{eq_phi2_2}) is reduced to the oscillator equation for $V(x)<E$
\begin{eqnarray}\label{eq_phi2_3}
\frac{d^2 \phi}{dx^2} + k^2(x) \phi &=& 0
\end{eqnarray}    
which can further be simplified at scales much smaller than the scale at which $V(x)$ is changing $k(x) \approx k$. As a result the approximate plane-wave solution is
$\phi(x) \approx \phi_0 e^{-i k(x). x} 
$. A three-simensional generalization of the one dimensional case should acquire the form
$\phi(\vec{r}) \approx \phi_0 e^{-i k(\vec{r}) \cdot \vec{r}} $.

Using this approximate plane wave solution we obtain for the probability density (both one and three dimensional cases)$
| \psi(t,\vec{r})|^2 = \chi_0^2  \phi_0^2 $, 
therefore
\begin{eqnarray}
\partial_t | \psi(t,\vec{r})|^2 &=& 0.
\end{eqnarray}    
The current density defined in (\ref{current j}) reduces to  $
j_x \approx -\frac{\hbar}{m} \left\{ k(x) + x \frac{d k}{dx} \right\}$
in the one dimensional case and $
\vec{j} \approx -\frac{\hbar}{m} [ \vec{k}(\vec{r})  + \nabla \vec{k}  \cdot \vec{r}]
$ for the three dimensional generalization. Note, the divergence of the current density is no longer a vanishing quantity
 \begin{eqnarray}
\frac{j_x}{d x}  \approx -\frac{\hbar}{m} [ 2\frac{d k}{dx} + \frac{d^2 k}{dx^2}]
\end{eqnarray}    
but rather large at step-like structures in the space-dependent potential. The three dimensional generalisation looks alike  
 \begin{eqnarray}
{\rm div } \vec{j} \approx -\frac{\hbar}{m} [ 4 \; {\rm div } \vec{k}  + (\Delta \vec{k}  \cdot \vec{r} ]
\end{eqnarray}    
and obviously the continuity equation $\partial_t | \psi|^2+ {\rm div} \vec{j} \neq 0$ no longer holds true, therefore { \it probability is not preserved}.

Contrary to that, energy is conserved. An evaluation of the energy of that state $
E_{{\rm state}}= i \hbar \int \psi^{\ast} \frac{\partial \psi}{\partial t} dr^3 = E , 
$ reveals that it is a constant and as a result {\it energy is conserved}
\begin{eqnarray}
\frac{ d E_{{\rm state}} }{d t} = 0.
\end{eqnarray}

\section{Conclusion}

In conclusion, let us put an arbitrary non-relativistic quantum system in a varying gravitational field such as the one produced by passing gravitational waves. The test quantum system will experience either the time-varying, space-varying or both space-time varying potential because of the interaction with the gravitational field. Usually the interaction term is $\propto R \psi$, where $R$ is the scalar curvature. In the case of $R(\vec{r})$ being space-dependent, gravitation will break the conservation of probability relation, which can be measured. As a result, by measuring the offset of the quantum state from obeying the continuity equation we will have a method of detecting the presence of space varying gravitational field. Interestingly, such sensitive macroscopic quantum systems exist in the form of superconducting or superfluid condensates.

Finally,  the mutual dynamic exclusion of probability and energy conservation opens up an opportunity to explore physical systems by producing ''equations of motion" which do not come from a variational principle applied on an energy functional. Such novel equations of motion may come as proper factorizations of the continuity equation. Examples include the Schr\"odinger and the Dirac equations, but probably this description of reality allows for more equations of this sort. Probably even the proper one for quantum gravity.



\begin{thebibliography}{77}

\bibitem{goldstein} H. Goldstein, Ch. Poole and J. Safko, {\it Classical Mechanics} (3th ed.),  Addison Wesley (2002).

\bibitem{batchelor}  G. K. Batchelor, {\it An Introduction to Fluid Dynamics}, Cambridge University Press (1967).

\bibitem{EM} E. M. Purcell, D.J. Morin, {\it Electricity and magnetism} (3rd ed.). Cambridge University Press  (2013).



\end{thebibliography}
\end{document}